# Elastic and viscous properties of nematic dimer CB7CB

Greta Babakhanova,[1,2] Zeinab Parsouzi,[3] Sathyanarayana Paladugu,[1] Hao Wang,[1,2] Yu. A. Nastishin,[1,4] Sergij V. Shiyanovskii,[1] Samuel Sprunt,[3,†] and Oleg D. Lavrentovich[1,2,3*]

[1]*Liquid Crystal Institute, Kent State University, Kent, OH 44242, USA.*
[2]*Chemical Physics Interdisciplinary Program, Kent State University, Kent, OH 44242, USA.*
[3]*Department of Physics, Kent State University, Kent, OH 44242, USA.*
[4]*Hetman Petro Sahaidachnyi National Army Academy, 32, Heroes of Maidan St., Lviv 79012, Ukraine*

**Abstract**

We present a comprehensive set of measurements of optical, dielectric, diamagnetic, elastic and viscous properties in the nematic (N) phase formed by a liquid crystalline dimer. The studied dimer, 1,7-bis-4-(4′-cyanobiphenyl) heptane (CB7CB), is composed of two rigid rod-like cyanobiphenyl segments connected by a flexible aliphatic link with seven methyl groups. CB7CB and other nematic dimers are of interest due to their tendency to adopt bent configurations and to form two states possessing a modulated nematic director structure, namely, the twist bend nematic, $N_{TB}$, and the oblique helicoidal cholesteric, $Ch_{OH}$, which occurs when the achiral dimer is doped with a chiral additive and exposed to an external electric or magnetic field. We characterize the material parameters as functions of temperature in the entire temperature range of the N phase, including the pre-transitional regions near the N-$N_{TB}$ and N-to-isotropic (I) transitions. The splay constant $K_{11}$ is determined by two direct and independent techniques, namely, detection of the Frederiks transition and measurement of director fluctuation amplitudes by dynamic light scattering (DLS). The bend $K_{33}$ and twist $K_{22}$ constants are measured by DLS. $K_{33}$ being the smallest of the three constants, shows a strong non-monotonous temperature dependence with a negative slope in both N-I and N-$N_{TB}$ pretransitional regions. The measured ratio $K_{11}/K_{22}$ is larger than 2 in the entire nematic temperature range. The orientational viscosities associated with splay, twist and bend fluctuations in the N phase are comparable to those of nematics formed by rod-like molecules. All three show strong temperature dependence, increasing sharply near the N – $N_{TB}$ transition.

† ssprunt@kent.edu
* Corresponding author: olavrent@kent.edu

## I. INTRODUCTION

Liquid crystals spectacularly illustrate how subtle variation in molecular structure leads to dramatic changes in the macroscopic properties of a material. Rigid rod-like molecules such as 4′-pentyl-4-cyanobiphenyl (5CB) are known to form the nematic (N) phase with a long range orientational order and no positional order; in the N phase, the rod-like molecules are on average aligned along a single direction, specified by a unit vector $\hat{\mathbf{n}}$ called the director. However, when two cyanobiphenyl moieties are linked into a single molecule by a flexible aliphatic chain with an odd number of methyl groups, as in the case of 1,7-bis-4-(4′-cyanobiphenyl) heptane (CB7CB), a new, lower temperature nematic phase, the so-called twist-bend nematic ($N_{TB}$), emerges [1-3]. This phase exhibits a uniform mass density but a spatially modulated and locally chiral director field. The director precesses helically on a cone with an extremely small period (pitch), measured by transmission electron microscopy (TEM) [2,3] to be about 8 nm. The nanoscale heliconical geometry is evidenced by characteristic textures of asymmetric Bouligand arches in freeze-fracture TEM observations [3] and by resonant X-ray scattering [4].

The heliconical structure of the $N_{TB}$ phase, which was theoretically predicted by Meyer [5], Dozov [6] and Memmer [7], is a result of the tendency of molecules to adopt spontaneous bent conformations; the twist is required to yield a spatially uniform bend. In Dozov's theory, the transition from the N to $N_{TB}$ is associated with a change of the sign of the bend elastic constant $K_{33}$ from positive to negative. A periodically modulated N phase of an alternative splay-bend type can also be realized as a combination of periodic bend and splay [5,6]. The relative stability is controlled by the ratio of the splay $K_{11}$ to twist $K_{22}$ elastic constants; the $N_{TB}$ variant is stable when $\frac{K_{11}}{K_{22}} > 2$ [6].

The dimer material CB7CB, in addition to $N_{TB}$, also exhibits another twist-bend structure, the so-called oblique helicoidal cholesteric, $Ch_{OH}$, when doped with a small amount of chiral additive and subjected to an externally applied electric [8,9] or magnetic [10] field. Here again, the experimental observations followed earlier theoretical predictions by Meyer [11] and de Gennes [12]. In an ordinary cholesteric, the molecules twist in a helical fashion with the director remaining perpendicular to the helical axis. In the $Ch_{OH}$ structure, the director twists while making an angle $\theta < \pi/2$ with the twist axis and the direction of the applied field. The tilt introduces bend in addition to twist. However, when the bend elastic constant is small, the elastic energy penalty for bend is compensated by the dielectric energy gain associated with the nonzero projection of the local director on the field direction. Geometrically, the director structure of the $Ch_{OH}$ is similar to that of the $N_{TB}$ phase, but the pitch $P$ of the $Ch_{OH}$ is typically much larger than the nano-scale pitch of the $N_{TB}$, since the molecules in the $Ch_{OH}$ can rotate around their long axes [13], while the local structure of $N_{TB}$ inhibits this rotation and is essentially biaxial. Both $\theta$ and $P$ in $Ch_{OH}$ are explicit functions of the applied field and the ratio $\frac{K_{33}}{K_{22}}$ [8,11]. The sensitivity of the pitch to the electric field makes it possible to realize electrically controlled selective reflection of light [9] and lasing [14] within broad spectral ranges, which are potentially useful effects for practical applications of $Ch_{OH}$. The important parameter to optimize in these applications is the ratio $\frac{K_{33}}{K_{22}}$, which depend on temperature. Measurements such as those reported here are crucial to explore the potential for this optimization.

The sensitivity of both $N_{TB}$ and $Ch_{OH}$ to the elastic constants of the corresponding material in its N state motivate the work reported here. In particular, we experimentally determine the temperature dependence of all three bulk elastic constants in the entire range of the nematic phase of the dimer CB7CB. These three constants are deduced from the dynamic light scattering (DLS) data and also, in the case of splay constant, from the Frederiks transition threshold in the electric field. The elastic properties of CB7CB have been explored in the past [15-19], but the data obtained by different groups differ from each other rather substantially.

To date, the most complete study has been presented by Yun et al [15] who determined the temperature dependencies of all three bulk elastic constants by an electro-optical technique. Yun et al [15] used a Frederiks transition in a planar cell to determine the splay elastic constant $K_{11}$ from the threshold of director deformations caused by an applied electric field. As the field increases, the initial pure splay mode of deformations is replaced by a mixed splay-bend distortion. By fitting the capacitance response of the cell with an analytical expression in which the elastic parameter is of the form $\kappa = (K_{33} - K_{11})/K_{11}$, one can extract the value of $K_{33}$ since $K_{11}$ is known. The method has been originally proposed for nematics formed by rod-like molecules [20], in which $K_{33}$ is significantly larger than $K_{11}$, so that the fitting parameter $\kappa$ is large and the fitting of extrapolated response is robust. In the case of dimers, however, it is expected that the largest elastic constant is $K_{11}$ while $K_{33}$ is the smallest [1,21-23]. Therefore, the low energy cost of bend distortions and small contribution of $K_{33}$ to the fitting parameter $\kappa$ makes the extrapolation approach less robust for the dimers. The ratio $\frac{K_{11}}{K_{22}}$ was found to be around 1.4, i.e., smaller than 2, near the N- $N_{TB}$ phase transition [15]. The result is somewhat surprising, since the inequality $\frac{K_{11}}{K_{22}} > 2$ represents a criterion of the formation of the $N_{TB}$ phase as opposed to a splay-bend phase. The very fact of the twist-bend deformations in the low-temperature nematic phase has been established in the case of CB7CB by the freeze-fracture transmission microscopy studies [3] and by the resonant carbon soft X-ray scattering [4]. Sebastian et al and Lopez et al [16,19] used the same extrapolation technique of splay Frederiks transition to determine $K_{11}$ and $K_{33}$; it was found that as the temperature decreases towards the N-$N_{TB}$ transition, $K_{11}$ monotonously increases while $K_{33}$ first increases and then decreases. The values of $K_{11}$ determined in [16,19] were somewhat higher (by $\approx 4\%$) than those in Ref. [15], while the values of $K_{33}$ in [16,19] were higher than those in Ref. [15] by approximately a factor of 3.

Qualitatively different results were presented by Parthasarathi et al [18] who reported that as the temperature is lowered towards the N-$N_{TB}$ transition in pure CB7CB, the bend constant $K_{33}$ increases rather than decreases. Therefore, the prior experimental results on the elastic constants of the nematic phase of CB7CB are rather controversial. In light of the importance of this material for understanding of $N_{TB}$ and $Ch_{OH}$ structures, the issue needs to be revisited.

In this work, we use direct and complementary techniques to determine all three elastic constants of CB7CB as well as other material parameters. The electro-optic version of the Frederiks effect is used only to find the splay elastic constant $K_{11}$ from the direct measurements of the threshold voltage needed to cause director distortions. To find the bend and twist constants and to independently determine the splay constant, we use dynamic light scattering (DLS). The two independently determined values of $K_{11}$ served as a test of reliability. Light is scattered at fluctuations of the director which is the local optic axis of the nematic. By designing a proper geometry of the experiment (polarizations, incident and scattering angles), one can separate contributions of different modes of deformations and, in particular, probe the deformations of pure bend. As we demonstrate in this paper, this extraction of bend is especially well suited for the dimeric materials in which $K_{33}$ is the smallest of all three bulk constants. Furthermore, besides the direct information about the elastic properties, the DLS data, with a proper calibration, also yield the values of orientational viscosities, corresponding to the relaxation dynamics of splay, bend, and the predominantly twist component of twist-bend deformations, as described by Majumdar et al [24] and Zhou et al [25]; these viscosities are presented as functions of temperature. Finally, this work presents results on other material properties of CB7CB, including refractive indices, birefringence, dielectric permittivities and diamagnetic anisotropy. The results are of importance in deepening our understanding of the material properties leading to the $N_{TB}$ and $Ch_{OH}$ twist-bend states and in optimization of electrically tunable selective reflection of light and lasing utilizing the $Ch_{OH}$ structure.

## II. EXPERIMENTAL DETAILS

### A. Chemical structure, phase diagram, and alignment

The chemical structure and the phase diagram of CB7CB are shown in Fig. 1. Phase characterization was performed upon cooling with the rate of $0.1\,°/\min$ using polarizing optical microscopy (POM). The temperature was controlled with an Instec HCS402 hostage and mK2000 temperature controller with a temperature stability of $0.01\,°C$. The CB7CB was synthesized following the procedures reported by Chen et al [2]. Planar alignment was promoted by spin-coating PI2555 (HD Microsystems) polyimide layer on indium tin oxide (ITO)-coated glass substrates. The uniform alignment was achieved by rubbing the substrates with a velvet cloth. The cells were assembled with anti-parallel arrangement of the rubbing direction. The cell gap $d$ was controlled by Micropearl glass spacers (mixed with NOA 71 UV glue), and measured using a Perkin Elmer UV/VIS Spectrometer Lambda 18. All the experimental cells were filled by capillary action in the I phase. Figure 2 shows the POM textures of the N and $N_{TB}$ phases in a planar cell ($d = 18.9\ \mu m$).

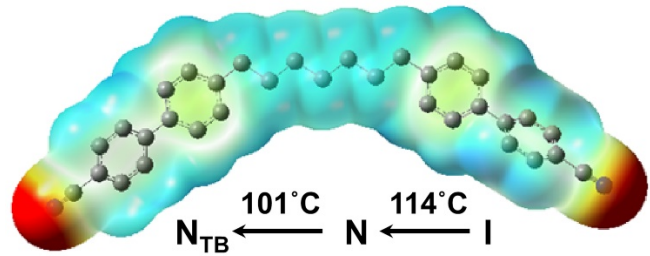

FIG 1. (Color online) Molecular structure of CB7CB with electrostatic potential surface and phase diagram upon cooling (negative and positive electric charges excesses are shown as red and blue, respectively).

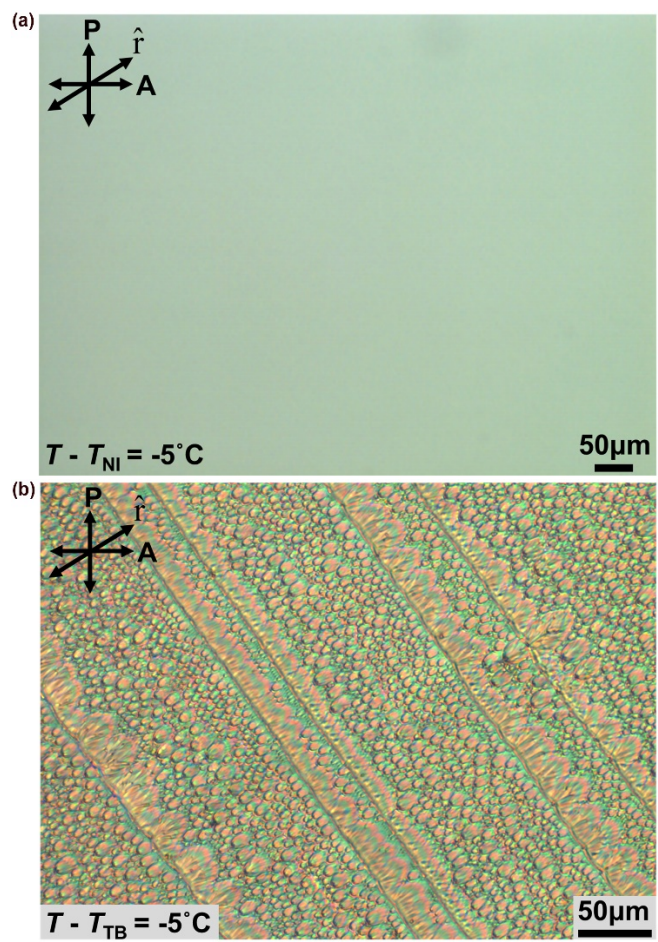

FIG 2. (Color online) POM textures of CB7CB under cross polarizers showing uniformly aligned (a) N and (b) $N_{TB}$ phases with focal conic domains. The direction of rubbing is shown by the axis $\hat{r}$.

### B. Refractive indices and birefringence

We used the wedge-cell technique [26] to determine the ordinary $n_o$ and extraordinary $n_e$ refractive indices of CB7CB. The wedge cell was prepared using planar, rubbed ITO substrates assembled in antiparallel fashion such that the rubbing direction was perpendicular to the wedge thickness gradient. The pretilt angle in the planar substrates measured using crystal rotation method [27] is less than $1°$. The thickness of the thick part of the wedge was set by a stripe of NOA 71 glue with pre-mixed Micropearl glass spacers. Initially, the optical interference technique described in [26] was used to determine the angle of an empty wedge, by shining a Helium-Neon (HeNe) laser beam ($\lambda = 633$ nm) onto the cell and recording an interference pattern under the POM. The temperature dependence of the wedge angle was determined over the same temperature range as the nematic range. The wedge cell was then filled with CB7CB by capillary action in the I phase. The analyzer (A) and polarizer (P) were aligned parallel to each other. Their orientations with respect to the nematic director were chosen to explore $n_o$ and $n_e$ independently. When polarization directions of both A and P are perpendicular or parallel to $\hat{\mathbf{n}}$, the multiple-beam interference in the wedge cell yielded $n_o$ or $n_e$, respectively, according to the following equation

$$n_{o,e} = \frac{l\lambda}{2\alpha(s_{m+l}^{o,e} - s_m^{o,e})}, \qquad (1)$$

where $l$ is the interference order (i.e. the fringe number), $\lambda$ is the wavelength of probing light, $\alpha$ is the wedge angle, and $(s_{m+l}^{o,e} - s_m^{o,e})$ is the distance between the interference maxima [26].

The Senarmont technique was employed to verify the birefringence, $\Delta n$, due to the higher sensitivity of this method [28]. The experimental optical set-up is displayed in Fig. 3. First, a polarizer and analyzer (positioned on motorized rotational stage), were crossed for maximum extinction. A quarter-wave plate was placed such that the optical axis is parallel to the initial polarizer. The planar cell ($d = 10.1\,\mu\text{m}$) was introduced with the rubbing direction, $\hat{r}$, making an angle $\varphi_o = 45°$ with the fist polarizer. The sample was probed with He-Ne laser light. For each temperature scan, the analyzer was rotated until the intensity of the linearly polarized light emerging from the quarter-wave plate reached a minimum ($I_{min}$) corresponding to an angle $\beta$. The maximum error of the angle measurements was $1°$. The total phase retardation was calculated as $\Delta\Phi = 2\beta + N2\pi$, where $N$ is an integer number, and the resultant birefringence is determined as

$$\Delta n = \frac{\lambda}{2\pi d}\Delta\Phi \quad (2)$$

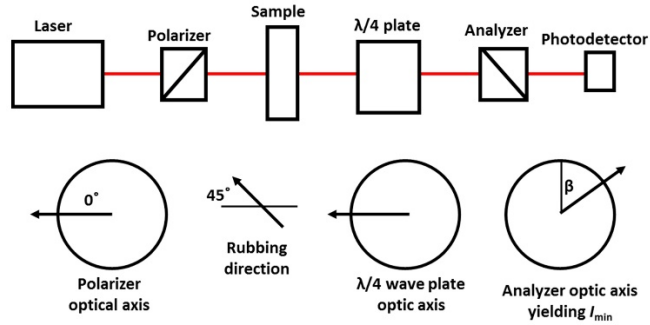

FIG 3. (Color online) Senarmont experimental set-up for birefringence measurements.

To diminish the contribution from experimental errors, we fitted the measured $\Delta n$ and $(n_e + n_o)$ data with a polynomial fit and obtained smoothed data for the temperature dependence for the refractive indices, according to the equation:

$$n_{e,o} = \frac{(n_e + n_o) \pm (n_e - n_o)}{2} \quad (3)$$

### C. Electro-optical measurements

Dielectric characterization was performed using a precision LCR meter 4284A (Hewlett Packard). Temperature-dependent dielectric permittivities were calculated from the capacitance measurements on a planar cell ($d = 18.9\,\mu\text{m}$) of CB7CB. A voltage ($V$) up to 20 $V_{rms}$ was applied across the active area of the patterned ITO electrodes. The square ITO patterned area was 25 mm². The measurements were performed at frequencies: $f = 5, 10, 20, 40$ and $60$ kHz. No significant dissipation was observed in the range 5-200 kHz. The perpendicular component of dielectric permittivity $\varepsilon_\perp$ was calculated from capacitance measurements at low voltages, below the Frederiks threshold, whereas the parallel component $\varepsilon_\parallel$ was determined by extrapolation method [29] at high voltages. The cell capacitance $C$ was plotted as a function of $V$ to find the splay Frederiks threshold voltage ($V_{th}$) by the double-line extrapolation method [29,30]. The splay elastic constant $K_{11}$ was then calculated according to

$$K_{11} = \frac{\varepsilon_o \Delta\varepsilon V_{th}^2}{\pi^2}, \qquad (4)$$

where $\Delta\varepsilon = \varepsilon_\parallel - \varepsilon_\perp$ and $\varepsilon_o$ is the vacuum permittivity.

### D. Diamagnetic anisotropy

The diamagnetic anisotropy, $\Delta\chi$, was determined utilizing a planar cell ($d = 18.9$ $\mu$m) placed in a uniform magnetic field applied perpendicular to the bounding plates (Fig. 4). The sample was positioned between two crossed polarizers with the rubbing direction $\hat{r}$ making an angle of $45°$ with the polarizer axes. The director reorientation caused by the splay Frederiks transition was monitored via transmitted light intensity data. Subsequently, the magnetic threshold ($B_{th}$) was extrapolated from measurements of the optical phase retardance vs. magnetic field curve using a double-line extrapolation approach [29,30]. The values of $\Delta\chi$ were calculated by relating electric and magnetic splay Frederiks effects according to

$$\Delta\chi = \varepsilon_o \mu_o \Delta\varepsilon \left(\frac{V_{th}}{dB_{th}}\right)^2 \qquad (5)$$

where $\mu_o = 4\pi \times 10^{-7}$ H/m is the vacuum permeability.

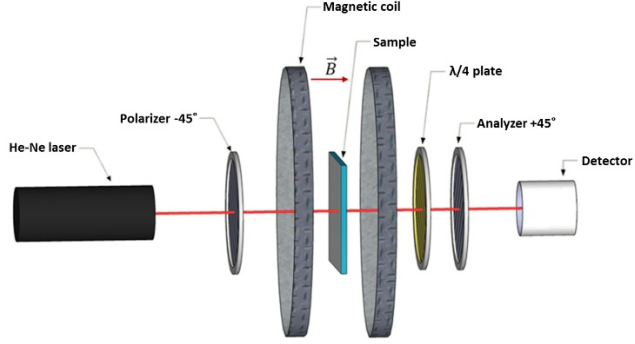

FIG 4. (Color online) Optical set-up to determine $\Delta\chi$.

**E. Dynamic light scattering**

Dynamic light scattering (DLS) on planar cells ($d = 16.5\ \mu$m) of CB7CB was conducted as a function of temperature on cooling. In order to obtain the absolute values of $K_{11}$, $K_{22}$ and $K_{33}$, we additionally recorded light scattering from a sample of the well-characterized calamitic nematic 4-*n*-octyloxy-4'-cyanobiphenyl (8OCB), for which the elastic constants and refractive indices are known with high precision [31-33]. The 8OCB experiments were performed at selected temperatures and under the same experimental conditions as for CB7CB, using a planar cell ($d = 14.4\ \mu$m). Specifically, 8OCB and CB7CB cells were situated in the same plane and placed adjacent to each other in the scattering apparatus. They could then be translated into or out of the incident laser beam by turning a single micrometer, with no other effect on the state or parameters of the experiment. High quality of homogeneous $\hat{\mathbf{n}}$ alignment was confirmed by POM, performed in situ on the DLS set-up. Phase transition temperatures of test samples were checked both before and after the experiment.

In the light scattering set-up, the output of a HeNe laser (Spectra-Physics, model 127), with a wavelength $\lambda_o = 633$ nm, incident power of 4 mW, and polarization oriented perpendicular to the scattering plane, is focused onto a spot on the sample with diameter ~50 μm. The hot stage containing the control and test samples was installed on a three-stage goniometer that allowed independent adjustment of the incident angle (set to normal incidence), the scattering angle, and the angle of the director $\hat{\mathbf{n}}$ with respect to the scattering plane. This flexibility enabled us to isolate scattering from three components of the director fluctuations:

pure bend ($K_{33}$), twist-bend mode dominated by twist ($K_{22}$), and pure splay ($K_{11}$) (Fig.5). In each case, depolarized scattered light was collected through a pinhole-lens-optical fiber detection layout. Scattered photons were converted to electronic pulses through a photomultiplier-amplifier-discriminator combination, allowing homodyne time correlation functions of the scattered intensity to be computed on a digital correlator.

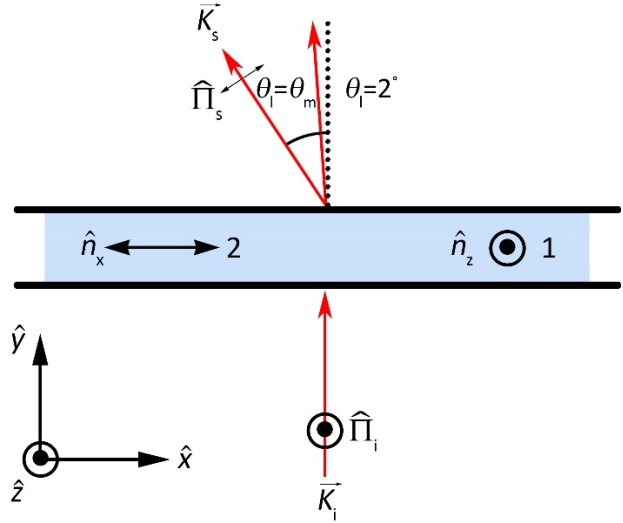

FIG 5. (Color online) Light scattering geometries. "1": splay+twist scattering; "2": pure bend scattering ($\theta_l = \theta_m$) and predominantly twist scattering ($\theta_l = 2°$). $\vec{K}_i$ and $\vec{K}_s$ correspond to incident and scattering wave vectors respectively; $\theta_l$ is scattering angle measured in the laboratory frame; $\theta_m$ refers to the so-called "magic angle"; $\Pi_i$ and $\Pi_s$ indicate the light of incident polarization and scattering polarizations which are orthogonal to each other.

In splay+twist geometry "1" of Fig. 5, the scattering vector $\vec{q}$ representing the difference between the incident and scattered wave-vectors, $\vec{q} = \vec{K}_s - \vec{K}_i$, is perpendicular to the director $\hat{\mathbf{n}}$ ($\vec{q} = \vec{q}_\perp$). In this case, splay and twist fluctuations contribute simultaneously to the scattering. The measured light intensity (divided by the incident light intensity $I_o$) for normal incidence is given by [34]:

$$\frac{I_{\mathrm{ST}}(\theta_1)}{I_\mathrm{o}} = (\Delta\varepsilon')^2(\pi\lambda^{-2})^2\Omega dAk_\mathrm{B}T\left[\frac{G_1(\theta_1)}{K_{11}q_\perp^2(\theta_1)} + \frac{G_2(\theta_1)}{K_{22}q_\perp^2(\theta_1)}\right], \qquad (6)$$

where $\theta_1$ is scattering angle in the laboratory, $\Delta\varepsilon'$ is optical dielectric anisotropy, $\lambda$ is the wavelength of light, $T$ is the absolute temperature, $\Omega$ is collection solid angle, $A$ is the cross-sectional area of the illuminated sample volume, and $d$ is the sample thickness. $G_1(\theta_1)$ and $G_2(\theta_1)$ are geometrical scattering factors dependent on $n_\perp, n_\parallel$ and $\theta_1$. By choosing the scattering angle to be the so-called "magic angle", $\theta_\mathrm{m} = \sin^{-1}\left(n_\perp\sqrt{1-\frac{n_\perp^2}{n_\parallel^2}}\right)$, the contribution of twist fluctuation mode to the total intensity disappears, so that $I_\mathrm{S}(\theta_\mathrm{m}) \propto \frac{G_1(\theta_\mathrm{m})}{K_{11}q_\perp^2(\theta_\mathrm{m})}$. In this case, we can extract the elastic constant $K_{11}$. The fitted $n_\perp$ and $n_\parallel$ values of CB7CB at the same wavelength as used for DLS were used to determine $\theta_\mathrm{m}$ at each temperature. The calculated values of $\theta_\mathrm{m}$ vary over $33°-40°$ within the N range. For the 8OCB control sample, we used literature values of the refractive indices to obtain $\theta_\mathrm{m}$ [35].

In geometry "2" ("twist-bend" geometry), where $\hat{\mathbf{n}}$ lies in scattering plane, the scattering comes from a combination of the twist-bend normal mode of director fluctuations [34]:

$$\frac{I_{\mathrm{TB}}(\theta_1)}{I_\mathrm{o}} = (\Delta\varepsilon')^2(\pi\lambda^{-2})^2\Omega dAk_\mathrm{B}T\left[\frac{G_3(\theta_1)}{K_{33}q_\parallel^2(\theta_1) + K_{22}q_\perp^2(\theta_1)}\right] \qquad (7)$$

Here $G_3(\theta_1)$ is a geometrical factor appropriate to the twist-bend geometry. At a low experimental scattering angle of $\theta_1 = 2°$, the ratio $\frac{q_\perp^2(\theta_1)}{q_\parallel^2(\theta_1)} \approx 15$. Since we also know that $K_{22} \ll K_{33}$ for dimer molecules with odd-numbered spacers [21-23], we conclude that $K_{22}q_\perp^2(\theta_1) \gg K_{33}q_\parallel^2(\theta_1)$ for $\theta_1 = 2°$, so that, to an excellent approximation, only twist fluctuations contribute in the above expression for the scattered intensity. On the other hand, when $\theta_1 = \theta_\mathrm{m}$, one finds that $q_\perp^2(\theta_\mathrm{m}) = 0$ [34], and in this case only bend fluctuations are probed. Thus, by switching from very small $\theta_1$ to $\theta_1 = \theta_m$, we can selectively probe the elastic constants $K_{22}$ and $K_{33}$.

As mentioned previously, in order to obtain the absolute values of the moduli, pure splay, bend, or twist scattering from CB7CB must be calibrated against the corresponding scattering from 8OCB. Literature values of optical parameters ($n_{\parallel,8OCB} = 1.65, n_{\perp,8OCB} = 1.50$) [33] and elastic constants ($K_{11,8OCB} = 5.5$ pN, $K_{22,8OCB} = 2.9$ pN and $K_{33,8OCB} = 6.05$ pN) [31,32] at $T - T_{NI} = -6\ °C$ were used in order to calculate intensity ratios between the 8OCB and CB7CB samples at fixed temperature ($T - T_{NI} = -9\ °C$). That enables a straightforward calculation of the elastic constants of CB7CB.

DLS also provides information on the orientational viscosities. At optical frequencies the director fluctuation modes are overdamped, and the standard expression for the homodyne intensity correlation function is [36],

$$\langle I(0,\theta_1)I(\tau,\theta_1)\rangle = I(\theta_1)^2 \left[1 + e^{-2\Gamma(\theta_1)\tau}\right] \quad (8)$$

where $\Gamma(\theta_1)$ is the relaxation rate of the fluctuations, and $\tau$ is the correlation delay time. Fits of the correlation data to this expression give the relaxation rates for the two director normal modes [34]:

$$\Gamma_\alpha(\theta_1) = \frac{(K_{33}q_\parallel^2 + K_\alpha q_\perp^2)}{\eta_\alpha(\vec{q})}, \quad \alpha = 1, 2 \quad (9)$$

The viscosities $\eta_\alpha(\vec{q})$ are combinations of the Leslie and Miesowicz viscosities of the nematic. From the scattering geometries described above, values of the elastic moduli, and the fitted relaxation rates, we calculate the orientational viscosities for pure bend and splay scattering geometries as [36]:

$$\eta_{splay} = \frac{K_{11}q_\perp^2}{\Gamma_1(\theta_m)} = \eta_1(q_\perp) = \gamma_1 - \frac{\alpha_3^2}{\eta_b} \quad (10)$$

$$\eta_{bend} = \frac{K_{33}q_\parallel^2}{\Gamma_2(\theta_m)} = \eta_2(q_\parallel) = \gamma_1 - \frac{\alpha_2^2}{\eta_c} \quad (11)$$

where $\gamma_1, \alpha_2, \alpha_3, \eta_b, \eta_c$ are fundamental viscosities of the nematic fluid discussed in standard texts [34].

The corresponding orientational diffusivities, $D$, are found using the following relationships [34]:

$$D_{splay} = \frac{K_{11}}{\eta_{splay}} \tag{12}$$

$$D_{bend} = \frac{K_{33}}{\eta_{bend}} \tag{13}$$

In the case of the twist-bend mode (geometry "2") for small scattering angle $\theta_1 \approx 2°$ and $\frac{q_\perp^2}{q_\parallel^2} \approx 15$, the orientational viscosity becomes [34]

$$\eta_{twist-bend} = \frac{K_{22} q_\perp^2}{\Gamma_2(\theta_1 = 2°)} = \eta_2(q_\perp^2 \approx 15 q_\parallel^2) \approx \gamma_1 - \frac{\alpha_2^2}{\eta_a} \frac{q_\parallel^2}{q_\perp^2} \approx \gamma_1 - \frac{\alpha_2^2}{15\eta_a} \tag{14}$$

In contrast to the situation with the elastic constants, the orientational viscosity $\eta_{twist-bend}$ cannot be reasonably approximated by the pure twist contribution ($\eta_{twist} = \gamma_1$), since the value of $\frac{\alpha_2^2}{\eta_a}$ is typically rather large. For example, using reported values of $\alpha_2$ and $\eta_a$ for the standard monomeric calamitic 5CB [37], we have $\frac{\alpha_2^2}{\eta_a} = 0.22$ Pa·s at a temperature 10°C below isotropic-nematic transition, which means that in 5CB the last term in Eq. (14) is 16% of the reported value of $\gamma_1 = 0.08$ Pa·s. Thus, in the following section we report the orientational viscosity in Eq. (14) and the corresponding diffusivity

$$D_{twist-bend} = \frac{K_{22}}{\eta_{twist-bend}} \tag{15}$$

### III. RESULTS

#### A. Refractive indices and birefringence

The measurements of the principal refractive indices and the birefringence at $\lambda = 633$ nm are presented in Fig.6.

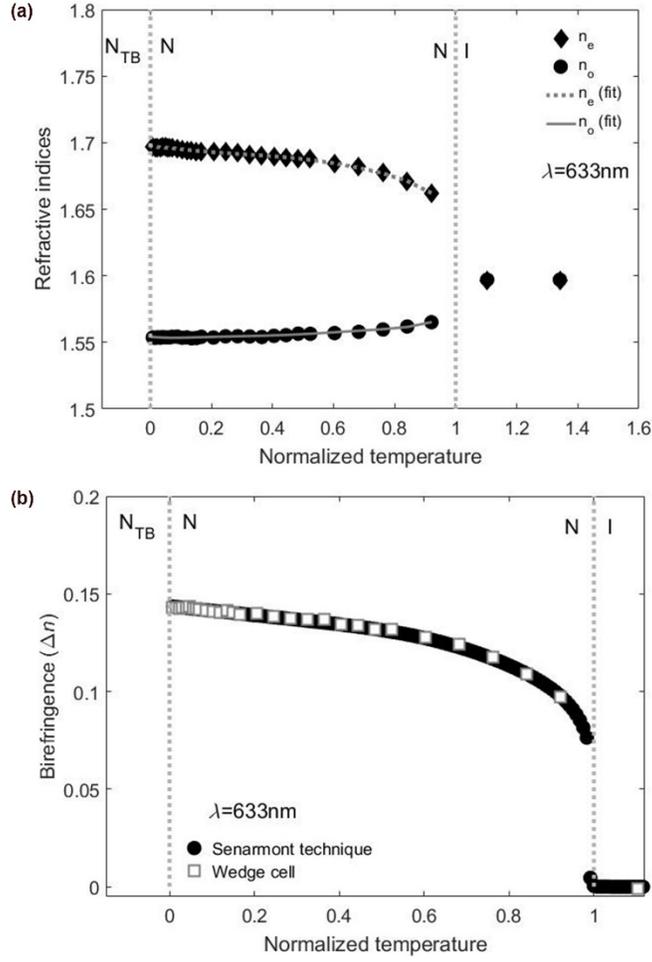

FIG 6. (Color online) (a) Refractive indices of CB7CB ($\lambda = 633$ nm); filled symbols represent the data acquired using a planar wedge cell; dotted and solid lines represent the fitted values of $n_o$ and $n_e$ respectively; (b) $\Delta n$ measured at $\lambda = 633$ nm using Senarmont technique (circles), planar wedge cell (squares).

The birefringence is positive and increases monotonically with decreasing temperature through the nematic phase down to the N-$N_{TB}$ transition.

### B. Dielectric and diamagnetic properties

Both components of dielectric permittivity ($\varepsilon_\parallel$ and $\varepsilon_\perp$) are presented in Fig.7(a) as functions of temperature. Fig.7(b) shows that, upon cooling from the I phase, $\Delta\varepsilon$ first significantly increases and then monotonously decreases. Changing the frequency in the range $f = 5-60$ kHz did not significantly affect the temperature dependent behavior of the dielectric constants.

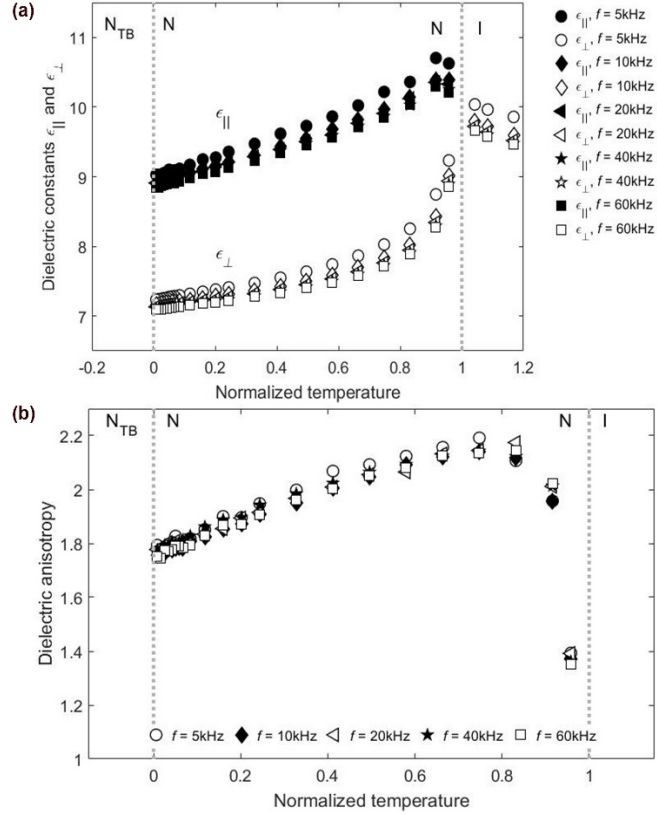

FIG 7. (a) Parallel and perpendicular components of temperature dependent dielectric constants and (b) $\Delta\varepsilon$ measured at $f = 5, 10, 20, 40$ and $60$ kHz using a planar cell ($d = 18.9\ \mu$m).

The diamagnetic anisotropy presented in Fig.8(a) has a sharp increase near the I-N transition. It grows with decreasing temperature and saturates towards the N-N$_{TB}$ transition. The data were fitted with a Haller's rule [38] of the form

$$\Delta\chi = \chi_\mathrm{o}\left(1 - \frac{T}{T^*}\right)^\nu \tag{16}$$

where $\chi_\mathrm{o} = 2.1493 \times 10^{-6}$, $T^* = 384.75$ and $\nu = 0.10953$ are the fitting parameters. The scalar nematic order parameter $S$ was calculated using the relationship $S = \frac{\Delta\chi(T)}{\chi_\mathrm{o}}$, and is presented in Fig.8(b).

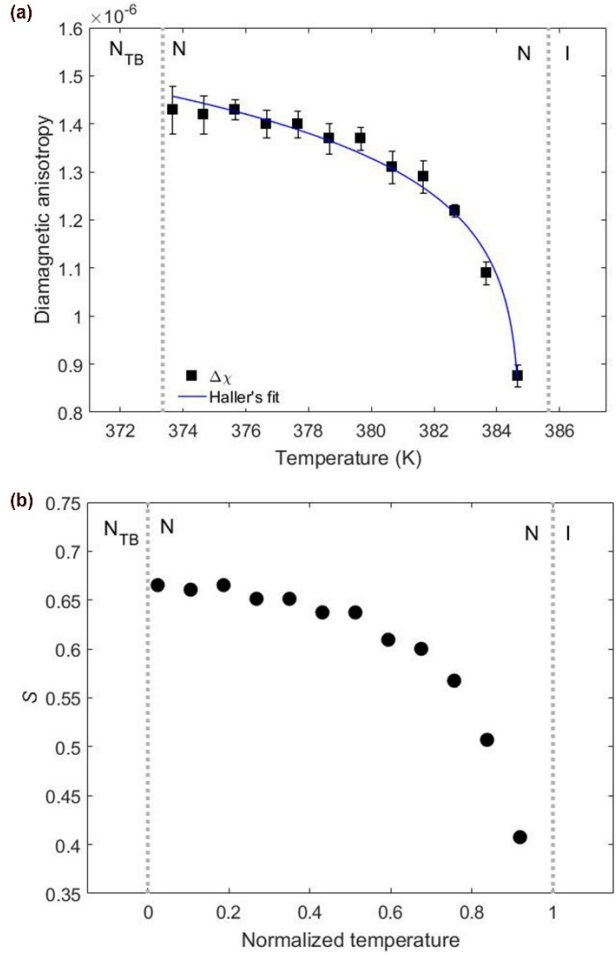

FIG 8. (Color online) (a) Temperature dependence of $\Delta\chi$ fitted with Haller's rule and (b) orientational order parameter.

**B. Elastic constants**

We performed direct measurements of all three elastic constants of CB7CB by the methods described above. The geometries described in Sec.II for DLS measurements allowed us to isolate the fluctuations related to splay, twist and bend. The independently calculated values of $K_{11}$, $K_{22}$ and $K_{33}$ from the intensity measurements are presented in Fig. 9(a). The Frederiks transition method was also employed to further assure the validity of DLS results. The results of capacitance measurements at $f = 60$ kHz yielding $K_{11}$ are presented in Fig. 9(a). On cooling,

$K_{11}$ weakly increases in the entire N range. There is also a slight increase in $K_{22}$, which is more prominent near the N-$N_{TB}$ transition. The most striking behavior, however, is seen in the development of $K_{33}$ as the temperature approaches the $N_{TB}$ phase, Fig. 9(b). On cooling through higher temperatures in the N phase $K_{33}$ increases slightly; however, on further cooling the value dramatically decreases approaching about 0.4 pN, after which there is a sharp growth near the N-$N_{TB}$ transition. The temperature-dependent elastic constant ratios, $\frac{K_{11}}{K_{22}}$, $\frac{K_{33}}{K_{22}}$ and $\frac{K_{11}}{K_{33}}$, are presented in Fig. 10.

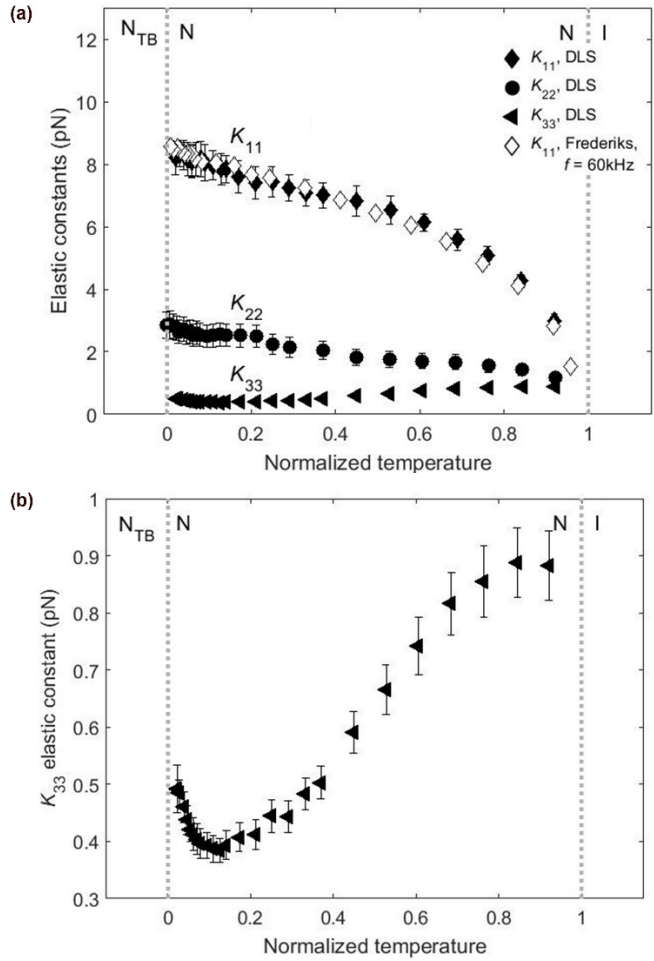

FIG 9. (Color online) (a) $K_{11}$, $K_{22}$ and $K_{33}$ measurements acquired using DLS measurements and $K_{11}$ data using capacitance method at $f = 60$ kHz; (b) temperature behavior of $K_{33}$ (DLS method).

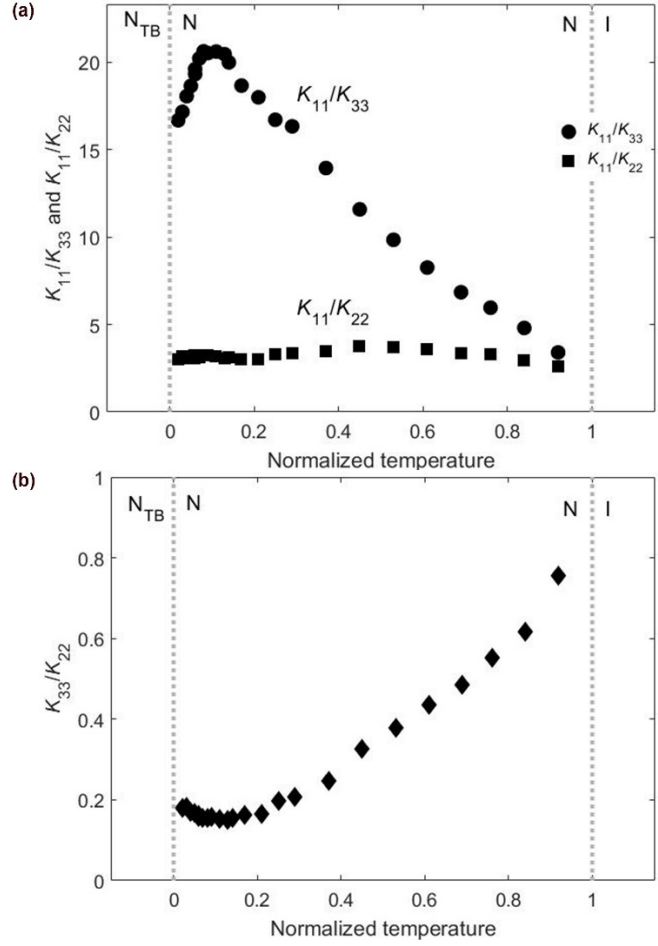

FIG 10. Ratio of elastic constants (a) $\dfrac{K_{11}}{K_{33}}$ and $\dfrac{K_{11}}{K_{22}}$, (b) $\dfrac{K_{33}}{K_{22}}$ from DLS data.

The temperature dependence of the three orientational viscosities, $\eta_{\text{splay}}$, $\eta_{\text{bend}}$, and $\eta_{\text{twist-bend}}$ are displayed in Fig. 11; they tend to increase upon cooling, especially on approaching the $N_{TB}$ phase. The corresponding diffusivities defined in Eqs. (12,13,15) are presented in Fig.12. As temperature is lowered from the I phase, $D_{\text{twist-bend}}$ remains practically constant, $D_{\text{splay}}$ gradually decreases, and $D_{\text{bend}}$ sharply decreases and levels off near the N-$N_{TB}$ transition.

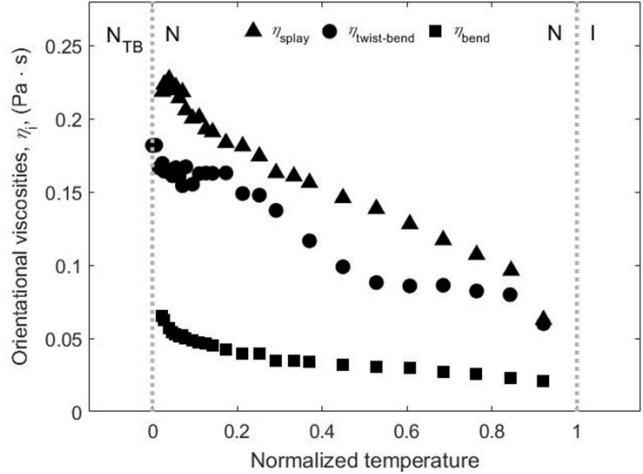

FIG 11. Temperature dependence of orientational viscosities.

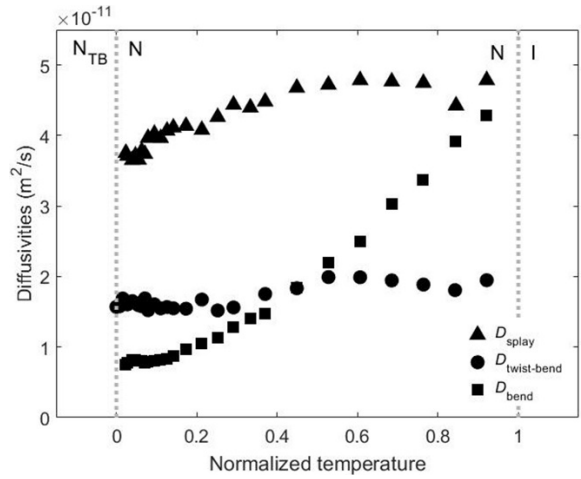

FIG 12. Temperature dependence of orientational diffusivities.

## IV. DISCUSSION

The behavior of the refractive indices and birefringence of CB7CB in the N phase is similar to that of ordinary calamatic liquid crystals such as 5CB. On cooling, the extraordinary refractive index increases and the ordinary index decreases. The birefringence is positive and grows with increasing orientational order. The birefringence results illustrated in Fig.6b are in good agreement with the most recent reports by Meyer et al (at $\lambda = 546$ nm) [39] and Tuchband et al (at $\lambda = 656.3$ nm) [40].

The tendency of the perpendicular component of the permittivity to decrease on cooling is similar to that for monomeric liquid crystals with positive $\Delta \varepsilon$ [1]. However, the parallel component of the dielectric permittivity does not behave as in a typical calamitic nematic liquid crystals with positive $\Delta\varepsilon$, where the parallel component is expected to increase on cooling. We show that $\varepsilon_\parallel$ increases slightly on cooling just below the I-N transition, similar to rod-like nematics, but upon further cooling rolls over and begins to decrease. As seen in Fig. 7, the decrease in $\varepsilon_\parallel$ leads to a decline in $\Delta\varepsilon$. Our experimental results are in good agreement with the reported dielectric studies of CB7CB and its longer homologues [1,15,16,18,23,41-43].

Dielectric permittivities are directly affected by molecular conformations [44]. In the case of the CB7CB, the molecular net dipole moment is determined mostly by the orientation of two cyanobiphenyl groups [1,16,43]. The flexible dimers respond to changes in temperature by modifying their conformations [1,45-48]. Theoretical models of flexible dimers suggest that there are two main populations of conformers, extended and hairpin-like, that vary in the angle between the two terminal cyanobiphenyl units [1]. This angle is about $120°$ for the extended conformers and $30°$ for the hairpin conformers. The population of extended conformers is growing as the order parameter increases upon cooling [1]. The extended conformers have a vanishing longitudinal dipole moment if the angle between two terminal groups exceeds $90°$ [16,43]. The effect of increased number of the extended conformers might contribute to the observed decrease of $\varepsilon_\parallel$ upon cooling in Fig.7a.

The diamagnetic anisotropy of CB7CB is positive, and, as expected, the temperature dependence is similar to that of the birefringence. The values of $\Delta\chi$ saturate near the N-N$_{TB}$ transition. The extrapolated values of the orientational order parameter using Haller's fit show that $S$ varies between 0.41 and 0.67, which is consistent with reported values [39,49].

The main focus of this report was to determine the elastic constants in the most accurate and complete fashion, which we achieved by direct measurements using both DLS and Frederiks transition methods. Previously, all three elastic constants were reported by Yun et al, however they expressed doubts about their $K_{22}$ measurements; moreover, literature values of $K_{33}$ have inconsistencies and were always fitted using splay Frederiks transition data [15,16,18,19]. In our study we employed direct measurements of all three elastic constants with very small temperature steps near N-N$_{TB}$ phase transition and show that $K_{11}$ values are in good agreement with the literature values that were obtained by capacitance method using the well-known

Frederiks transition technique [15,16,18,19]. Our values of $K_{22}$, however, are lower than the values reported in Ref. [15] by a factor of $\square 1.7$ near N-N$_{TB}$ transition. The ratio $\frac{K_{11}}{K_{22}}$ measured in our work is consistently higher than 2.5 in the entire N range of CB7CB, while Ref. [15] reports that on average $\frac{K_{11}}{K_{22}} \approx 1.4$, which is rather questionable since the twist-bend phase is predicted to be stable only when $\frac{K_{11}}{K_{22}} > 2$, as opposed to the splay-bend case in which one expects $\frac{K_{11}}{K_{22}} < 2$ [6,50].

Our measurements of $K_{33}$ show an overall decrease upon cooling to $\approx 0.38$ pN near the N-N$_{TB}$ transition, after which $K_{33}$ grows just before the transition. A slight increase in $K_{33}$ was already observed for CB7CB and other dimeric liquid crystals [16,19,21-23,43], with the exception of Ref. [18] that shows a monotonously increasing $K_{33}$ with the decreasing temperature.

## V. CONCLUSION

In this report, we experimentally determined the temperature-dependent optical, dielectric, diamagnetic and viscoelastic material properties of the most studied odd methylene-linked dimer, CB7CB, in its nematic phase. We determined that on cooling, the refractive indices behave much like the regular calamatic liquid crystals such as 5CB. The dielectric measurements reveal that the parallel component of dielectric permittivity decreases on approach to the N-N$_{TB}$ transition after an initial increase below the isotropic phase, while the perpendicular component decreases monotonically. The resultant dielectric anisotropy has a sharp increase below the clearing temperature, but then rolls over and gradually decreases on approaching the N-N$_{TB}$ transition. These results are consistent with the dimers adopting a more bent (less extended) average conformation as the N-N$_{TB}$ transition is approached.

We also reported the first independent temperature dependent measurements of all three elastic constants for CB7CB. We show that the relationship $K_{11} > K_{22} > K_{33}$ holds true in the entire nematic range, with $\frac{K_{11}}{K_{22}} > 2$ as predicted by the theory for a system that exhibits the $N_{TB}$ phase [6,50]. $K_{11}$ increases monotonously; $K_{22}$ also increases gradually but has a more prominent pretransitional increase close to the N-$N_{TB}$ phase transition. The temperature dependence of $K_{33}$ is the most dramatic – on cooling from the I-N transition it decreases to an unusually low value $\approx 0.38$ pN and then experiences pretransitional increase near the N-$N_{TB}$ transition. The pretransitional increase of both $K_{22}$ and $K_{33}$ might be explained by formation of clusters with periodic twist-bend modulation of the director close to $N_{TB}$ phase transition. These clusters lead to an increase of $K_{22}$ and $K_{33}$ since the equidistance of pseudolayers hinders twist and bend deformations of the director. A similar pretransitional increase of $K_{22}$ and $K_{33}$ is well documented near the nematic-to-smectic A phase transition [32,51,52]. The ratio $\frac{K_{33}}{K_{22}}$ that controls the field-tunable range of the heliconical structure [53] is less than 1 in the entire N range, thus making CB7CB a good candidate for preparation of heliconical cholesterics with a tunable pitch.

Finally, the calculated values of all three orientational viscosities increase on cooling, showing steeper slopes near the N-$N_{TB}$ phase transition. The bend orientational diffusivity sharply decreases on cooling from $T_{NI}$ and saturates near $T_{N_{TB}}$. The splay diffusivity decreases smoothly, whereas, on average, the twist diffusivity remains constant throughout the nematic range.

## Acknowledgments

The work was supported by NSF grants DMR-1410378 and DMR-1307674. The authors acknowledge R. Stayshich for purifying 8OCB.

G.B. and Z.P. contributed equally to this work.